\documentclass[12pt,psfig]{article}
\usepackage{amssymb}
\usepackage{amsmath}
\usepackage{epsfig}
\makeatletter
\def\@cite#1#2{{[{#1}]\if@tempswa\typeout
{IJCGA warning: optional citation argument
ignored: `#2'} \fi}}

\newcount\@tempcntc
\def\@citex[#1]#2{\if@filesw\immediate\write\@auxout{\string\citation{#2}}\fi
  \@tempcnta\z@\@tempcntb\m@ne\def\@citea{}\@cite{\@for\@citeb:=#2\do
    {\@ifundefined
       {b@\@citeb}{\@citeo\@tempcntb\m@ne\@citea\def\@citea{,}{\bf ?}\@warning
       {Citation `\@citeb' on page \thepage \space undefined}}%
    {\setbox\z@\hbox{\global\@tempcntc0\csname b@\@citeb\endcsname\relax}%
     \ifnum\@tempcntc=\z@ \@citeo\@tempcntb\m@ne
       \@citea\def\@citea{,}\hbox{\csname b@\@citeb\endcsname}%
     \else
      \advance\@tempcntb\@ne
      \ifnum\@tempcntb=\@tempcntc
      \else\advance\@tempcntb\m@ne\@citeo
      \@tempcnta\@tempcntc\@tempcntb\@tempcntc\fi\fi}}\@citeo}{#1}}
\def\@citeo{\ifnum\@tempcnta>\@tempcntb\else\@citea\def\@citea{,}%
  \ifnum\@tempcnta=\@tempcntb\the\@tempcnta\else
   {\advance\@tempcnta\@ne\ifnum\@tempcnta=\@tempcntb \else
\def\@citea{--}\fi
    \advance\@tempcnta\m@ne\the\@tempcnta\@citea\the\@tempcntb}\fi\fi}
\makeatother

\newcommand{\gsim}{\lower.7ex\hbox{$\;\stackrel{\textstyle>}{\sim}\;$}}
\newcommand{\lsim}{\lower.7ex\hbox{$\;\stackrel{\textstyle<}{\sim}\;$}}
\newcommand{\be}{\begin{equation}}
\newcommand{\ee}{\end{equation}}
\newcommand{\bea}{\begin{eqnarray}}
\newcommand{\eea}{\end{eqnarray}}
\newcommand{\f}{\frac}

\usepackage{amssymb}

\def\baselinestretch{1}
\begin{document}
\catcode`@=11
\newtoks\@stequation
\def\subequations{\refstepcounter{equation}%
\edef\@savedequation{\the\c@equation}%
  \@stequation=\expandafter{\theequation}
  \edef\@savedtheequation{\the\@stequation}
  \edef\oldtheequation{\theequation}%
  \setcounter{equation}{0}%
  \def\theequation{\oldtheequation\alph{equation}}}
\def\endsubequations{\setcounter{equation}{\@savedequation}%
  \@stequation=\expandafter{\@savedtheequation}%
  \edef\theequation{\the\@stequation}\global\@ignoretrue

\noindent}
\catcode`@=12
\begin{titlepage}
\title{{\bf
$f(R)$ actions, cosmic acceleration and local tests of gravity
}}
\vskip2in

\author{
{\bf Ignacio Navarro$$\footnote{\baselineskip=16pt E-mail: {\tt
i.navarro@damtp.cam.ac.uk}}} $\;\;$and$\;\;$ {\bf Karel Van
Acoleyen$$\footnote{\baselineskip=16pt E-mail: {\tt
k.van-acoleyen@damtp.cam.ac.uk}}}
\hspace{3cm}\\
 $$ {\small DAMTP, University of Cambridge, CB3 0WA Cambridge, UK}.
}

\date{}
\maketitle
\def\baselinestretch{1.15}
\begin{abstract}
\noindent

We study spherically symmetric solutions in $f(R)$ theories and its
compatibility with local tests of gravity. We start by clarifying
the range of validity of the weak field expansion and show that for
many models proposed to address the Dark Energy problem this
expansion breaks down in realistic situations. This invalidates the
conclusions of several papers that make inappropriate use of this
expansion. For the stable models that modify gravity only at small
curvatures we find that when the asymptotic background curvature is
large we approximately recover the solutions of Einstein gravity
through the so-called Chameleon mechanism, as a result of the
non-linear dynamics of the extra scalar degree of freedom contained
in the metric. In these models one would observe a transition from
Einstein to scalar-tensor gravity as the Universe expands and the
background curvature diminishes. Assuming an adiabatic evolution we
estimate the redshift at which this transition would take place for
a source with given mass and radius. We also show that models of
dynamical Dark Energy claimed to be compatible with tests of gravity
because the mass of the scalar is large in vacuum ($e.g.$ those that
also include $R^2$ corrections in the action), are not viable.

\end{abstract}

\vskip-19cm \rightline{} \rightline{DAMTP-2006-115} \vskip3in

\end{titlepage}
\setcounter{footnote}{0} \setcounter{page}{1}
\newpage
\baselineskip=20pt





\section{Introduction}

Theories of gravity whose action is some function of the Ricci
scalar have received much attention recently as possible models of
dynamical Dark Energy. It is well known that when we take a
gravitational Lagrangian as ${\cal L}_g=M_p^2 f(R)/2$ (where $M_p$
is the Planck mass and $R$ the scalar curvature), as long as one can
invert the relation \be e^{\sqrt{\frac{2}{3}}\f{\phi}{M_p}} =
\frac{d f(R)}{d R}\label{inv} \ee to find $R(\phi)$ for a range of
$R$, a solution where $R$ lies in this range can also be found from
an equivalent action that consists of Einstein gravity minimally
coupled to a scalar field, namely $\phi$,  with certain potential
\cite{Whitt:1984pd,Chiba:2003ir}. So these models, when applied to
the Dark Energy problem, are equivalent to quintessence models where
the scalar field is conformally coupled to matter\footnote{In this
paper we restrict ourselves to the conventional metric formulation
of $f(R)$ gravity and do not consider the Palatini formulation
\cite{Vollick:2003aw}.}. It has been shown that for many choices of
the function $f$ one obtains acceptable cosmological solutions that
could be used to describe the acceleration of the universe: in fact
one can reproduce an arbitrary expansion history by choosing $f$
suitably \cite{Capozziello:2005ku,Song:2006ej}. What has been
however subject of controversy is whether one can, at the same time
that one describes the Dark Energy in this way, satisfy the bounds
on the existence of extra scalar fields with gravitational couplings
coming from Solar System or laboratory experiments (see $e.g.$
\cite{Chiba:2003ir,Olmo:2005zr,Erickcek:2006vf,Cembranos:2005fi}).

Let us explain here the reasons why this controversy has arisen.
In order to check the experimental implications of a gravitational
theory for weak fields one usually starts by considering a
background solution and making an expansion of the equations of
motion (EOM) in powers of the fluctuations over this background
solution. If the second (and higher) order terms in this expansion
are negligible with respect to the first order ones in a
considered solution, perturbation theory is applicable. When this
is the case, the EOM can be approximated by a set of linear second
order differential equations, and we can usually find analytic
solutions for static and symmetric sources and check the
compatibility with experiments. The situation is however
completely different when this expansion is not applicable. In
this case we have to deal with non-linear differential equations
that are usually difficult to solve, and addressing the issue of
compatibility with experiments becomes challenging. We will see
that this is indeed the situation for many of the considered
$f(R)$ gravities: perturbation theory breaks down in some cases
and is not applicable because non-linearities in the equations are
not negligible.

In \cite{Chiba:2003ir,Olmo:2005zr,Erickcek:2006vf} this linearized
expansion was considered for some $f(R)$ theories. As we said this theory
generically consists of Einstein gravity plus an extra scalar
field conformally coupled to matter, so at the linear level the
only terms that appear in the EOM for the scalar are the kinetic
and mass terms with a source given by the trace of the
energy-momentum tensor (EMT). It was therefore argued in
\cite{Chiba:2003ir,Olmo:2005zr,Erickcek:2006vf} that simple models
in which the mass of the scalar is very small in vacuum, such as
$f=R\pm \mu^{2n+2}/R^n$, are ruled out because it would mediate an
extra long range force contradicting experiments\footnote{In these
cases the mass of the scalar in vacuum would be $\sim
  \mu$, and the application of these theories to
  the Dark energy problem demands $\mu \sim H_0$, where $H_0$ is the
value of the Hubble constant today.}. But this analysis was
challenged in \cite{Cembranos:2005fi} where solutions with $R \gg
\mu^2$ where considered. Intuitively, we can expect by looking at
the action above that when $R\gg \mu^2$ the effect of the
modification should be negligible. The extra terms in the equations
will be suppressed by powers of the ratio $\mu^2/R$ and the solution
for any given source should be very close to that of General
Relativity (GR) up to these small corrections. But while this is
true, we will see in this paper that in fact there is no
contradiction with the previous analysis. This is so because, in the
models with inverse powers of $R$, whenever the scalar curvature
deviates from its vacuum value significantly in some region ($\Delta R/R_0 \geq 1$
with $R_0\sim \mu^2$), the linearized expansion breaks down, and the analysis of
\cite{Chiba:2003ir,Olmo:2005zr,Erickcek:2006vf} can not be applied
to those solutions. It is then conceivable that in these models an
effect similar to the Chameleon mechanism of \cite{Khoury:2003aq}
could take place, where the effects of a scalar field that is very
light in vacuum are hidden because non-linearities are important
when considering sources. As we will see, the asymptotic boundary
conditions for the scalar curvature are the crucial element that
will allow us to decide whether there is a Chameleon effect or not
for the extra scalar degree of freedom. In particular, for the
models with inverse powers of $R$, if the background curvature goes
to a small value asymptotically ($R_0\sim \mu^2$), the Chameleon
effect does not take place: astrophysical bodies do not have a
``thin shell'' for static solutions. In this case the perturbative
solutions of \cite{Chiba:2003ir,Olmo:2005zr,Erickcek:2006vf} are
valid and the scalar curvature is ``locked'' to the background value
$everywhere$, $R\sim R_0 \sim \mu^2$ (a fact also noticed in
\cite{Erickcek:2006vf}). However, when the asymptotic value of the
curvature is $R_0\gg \mu^2$, the Chameleon effect $does$ take place
and we recover the solutions of Einstein gravity up to small
corrections\footnote{We are assuming here that the mass squared of
the scalar is positive in this regime, so the solutions with $R\gg
\mu^2$ are stable. For instance, when $f=R\pm \mu^{2n+2}/R^n$ this
would be the case only for the positive sign, see
\cite{Dolgov:2003px}. We will come back to this issue in section
four.}. We get then the following picture of gravitational dynamics
in these theories when applied to the Universe: one would observe a
transition from Einstein gravity at early times (when $H^2 \gg
\mu^2$) to scalar-tensor gravity at late times (when $H^2 \sim
\mu^2$). If we assume that this transition is adiabatic, $i.e.$ the
solution is always taken to be the equilibrium one, these types of
modification are ruled out, since we would be in the scalar-tensor
regime at present. However, in this paper we do no not rule out the
possibility that there are some models for which this transition is
non-adiabatic and slow enough so we would still remain in the ``GR
regime" locally in the Solar System. If this was the case those
models would not be in conflict with local tests of gravity. The
Chameleon effect will also allow us to resolve the apparent
discontinuity in the General Relativistic $\mu\rightarrow 0$ limit
of these theories where we seem to end up with a massless scalar
field coupled to gravity instead of GR. In the $real$ Universe
($i.e$ with a non-zero background energy density) we would recover
the usual gravitational dynamics in this limit only at a
non-perturbative level.

On the other hand, it has been shown that for some specific forms of
the action the scalar field can be made arbitrarily massive in
vacuum \cite{Dick:2003dw,Nojiri:2003ft}. If this mass is large
enough it is claimed that the scalar would have passed undetected so
the models do not conflict with local tests of gravity. But as we
will see this conclusion is wrong and this possibility can not be
realized. The reason for this is that we can not trust a weak field
expansion in this case: as we raise the mass of the scalar field, we
lower the energy scale where non-linearities become important. We
will find that for most functions $f$ that attempt to describe Dark
Energy the energy scale at which non-linearities become relevant is
$\Lambda_s \simeq M_p \left(\frac{H_0}{m_s}\right)^4$, where $m_s$
is the mass of the scalar field\footnote{And in all models that
satisfy some ``minimal requirements'' to be defined precisely later,
it is at least of order $\Lambda_s \simeq M_p
\left(\frac{H_0}{m_s}\right)^2$.}. If this mass is large compared
with $H_0$, as it should be in order to avoid conflict with
experiment, this scale is very small which means that in all
realistic situations where we want to apply the linearization over
vacuum it is not possible to do so because non-linearities in the
equations are not negligible. So in this case the fact that the mass
term appearing in the linearized EOM for the scalar in vacuum is
large does not imply compatibility with experiments. And we will
argue that, for observationally relevant distances, the true solution on vacuum
is again of the scalar-tensor type with an effectively massless scalar, as
in the generic models. Furthermore, we will show that models that
include positive powers of $R$ to raise the scalar mass in vacuum will
also conflict with experimental results on large curvature
backgrounds, where $R\gg \mu^2$. In those backgrounds the mass of the scalar
diminishes to a small value and the strong coupling scale
$\Lambda_s$ raises so one can use the linearized equations and it
is easy to conclude that one should have observed the effects
of the scalar on laboratory searches of fifth forces, for instance.

The paper is organized as follows. In the next section we will
review the solutions of the linearized EOM for a spherically
symmetric mass in $f(R)$ gravity and we will discuss the range of
validity of the weak field expansion for a generic action $f(R)$.
In the third section we will focus on theories designed
to explain the cosmic acceleration and we will consider the
expansion over vacuum where $R_0 \sim \mu^2 \sim H_0^2$. We will
distinguish two cases: the ``generic'' one (in which the scalar mass
is $\sim \mu$) and the ``fine-tuned'' one (in which the mass is $\gg
\mu$). In the first case we will comment on the curious property
that the scalar curvature remains small in the perturbative
solutions even in those regions where there is a large
energy-momentum density. In this case the strong-coupling scale of
the linearization is the Planck mass, so once the curvature gets
locked into this phase it remains there: one would require
Planck scale energies or very strong gravitational fields in order
to excite the scalar curvature out of its vacuum value. In the
second case we will explain why when we raise the mass of the scalar
in a certain background in a model of dynamical Dark Energy, the
linearized expansion over such background is actually of no use. And
we will argue that the true solution exhibits the same behavior as for
the generic case. Furthermore we will explain why the addition of
positive powers of $R$ to the action to yield this effect in vacuum
would imply a conflict with experiment whenever $R\gg \mu^2$. In the
fourth section we will briefly discuss the Chameleon mechanism and
the recovery of Einstein gravity in models that modify gravity only
at small curvatures $\sim \mu^2$, as long as the asymptotic
background curvature is $R_0 \gg \mu^2$ and the scalar mass squared
is positive. We will also discuss the transition from GR to
scalar-tensor gravity (or ``locked curvature'' phase) that would
take place in these theories when applied to the Universe as the
background curvature diminishes to a value $\sim \mu^2$ and the
Chameleon effect gradually disappears. Finally we end with the
conclusions in section five.

\section{The short distance weak field expansion, general case}

It is well known that the EOM derived from the action  \be S=
\f{1}{16\pi G_N}\int d^4x\sqrt{-g} f(R)+S_m\,\,\,\,\,
(\textrm{with}\,\,G_N\equiv 1/(8\pi M_p^2))\,,\label{action}\ee
where $S_m$ is the matter action minimally coupled to the metric,
are equivalent to those of Einstein gravity minimally coupled to a
scalar field that is conformally coupled to matter. This equivalence
is most easily demonstrated directly at the level of the action
\cite{Whitt:1984pd,Chiba:2003ir}, but it is also instructive to take
a slight detour and show how the equivalence arises at the level of
the EOM. The EOM derived from (\ref{action}) read: \be
E_{\mu\nu}(g_{\mu\nu})=f'(R)R_{\mu\nu}
-f(R)\f{g_{\mu\nu}}{2}+g_{\mu\nu}\Box f'(R)-\nabla_{\mu}\nabla_{\nu}
 f'(R)=8\pi G_N T_{\mu\nu}\,. \label{eom}\ee
A striking difference with Einstein gravity is seen from the
trace: \be E_{\mu}^{\mu}(g_{\mu\nu})= -2f(R)+f'(R)R+3\Box
f'(R)=8\pi G_N T\,.\ee In contrast with GR, where the Ricci scalar
is completely determined by the trace of the energy-momentum
tensor ($R=-8\pi G_N T$) we now see that $R$, or rather $f'(R)$,
is an independent propagating degree of freedom. It will therefore
be convenient to introduce a new variable $\psi$, that we will
identify with $f'(R)$, and cast the original set of fourth order
differential equations in $g_{\mu\nu}$ into a set of second order
equations in $g_{\mu\nu}$ and $\psi$: \bea
E_{\mu\nu}(g_{\mu\nu},\psi)&\equiv&\psi R_{\mu\nu}
-f(\mathcal{R}(\psi))\f{g_{\mu\nu}}{2}+g_{\mu\nu}\Box
\psi-\nabla_{\mu}\nabla_{\nu}
 \psi=8\pi G_N T_{\mu\nu}\,, \label{eomgpsi} \\
\mathcal{R}(\psi)&=&R \label{defpsi}\,,\eea
 where $\mathcal{R}$ is defined as the inverse of $f'$: \be f'(\mathcal{R}(\psi))
 \equiv\psi\,.\label{defpsi2}\ee It is clear that a solution of eqs.(\ref{eomgpsi},\ref{defpsi}) also
 solves the original EOM (\ref{eom}) as long as $\mathcal{R}(\psi)$ is
 well defined on the solution. Notice that the invertibility of $f^\prime$ is equivalent to the
 more familiar condition that $f''(R)$ should be
 different from zero. In general $f''(R)$ can become zero for some
 values of $R$ so $\mathcal{R}$ will be a
 multivalued function, and one has to chose a certain branch. The
 solutions of (\ref{eom}) are found then by considering
 eqs.(\ref{eomgpsi},\ref{defpsi}) with all possible branches.

 Finally, it turns out to be convenient to define a new metric variable \be \hat{g}_{\mu\nu}\equiv
f'(R)g_{\mu\nu}=\psi g_{\mu\nu}\,,\ee since in terms of this metric
the eqs.(\ref{eomgpsi},\ref{defpsi}) let themselves reshuffle as those
of Einstein gravity, minimally coupled to the scalar $\psi$ and with
a metric $g_{\mu\nu}=\hat{g}_{\mu\nu}/\psi$ in the matter sector:
\bea \hat{E}_{\mu\nu}&\equiv&
\psi^{-1}E_{\mu\nu}(g_{\mu\nu},\psi)+\f{g_{\mu\nu}}{2}(\mathcal{R}(\psi)-R)\nonumber\\
&=&\hat{G}_{\mu\nu}-\f{3}{2\psi^2}\partial_\mu\psi\partial_\nu\psi
+\hat{g}_{\mu\nu}\left(\f{3}{4\psi^2}\hat{g}^{\alpha\beta}\partial_\alpha\psi\partial_\beta\psi
+\f{1}{2\psi}\left(\mathcal{R}(\psi)-\f{f(\mathcal{R}(\psi))}{\psi}\right)\right)\,,\label{eqhat}\nonumber\\
&=&8\pi \f{G_N}{\psi} T_{\mu\nu}\,,\\
\nonumber\\
E_\psi&\equiv&E_{\mu}^{\mu}(g_{\mu\nu},\psi)+\psi(\mathcal{R}(\psi)-R)=
3\Box \psi+\mathcal{V}_p(\psi)=8\pi G_N T\,,\label{trace}\eea where
$\hat{G}_{\mu\nu}$ is the Einstein tensor corresponding to
$\hat{g}_{\mu\nu}$ and we have defined \be \mathcal{V}_p(\psi)\equiv
\psi\mathcal{R}(\psi)-2f(\mathcal{R}(\psi))\,. \label{defF}\ee  One
can now verify that these equations indeed arise from the variation with
respect to $\hat g_{\mu\nu}$ and $\psi$ of the action \bea
S_{eq}&\equiv& \f{1}{16\pi G_N}\int
d^4x\sqrt{-\hat{g}}\left(\hat{R}-\f{3}{2\psi^2}
\partial_{\mu}\psi\partial_{\nu}\psi\hat{g}^{\mu\nu}-\mathcal{V}(\psi)\right)
+S_m(\hat{g}_{\mu\nu}/\psi)\,,\nonumber\\
&=&\int
d^4x\sqrt{-\hat{g}}\left(\f{M_p^2}{2}\hat{R}-\f{1}{2}\partial_{\mu}\phi\partial_{\nu}\phi\hat{g}^{\mu\nu}
-V(\phi)\right)+S_m(\hat{g}_{\mu\nu}e^{-\kappa
\phi})\,,\label{action2}\eea where \be
\mathcal{V}(\psi)=\f{1}{\psi}(\mathcal{R}(\psi)-\f{f(\mathcal{R}(\psi))}{\psi})\,,
\ee and in the last line we have reparameterized the scalar: \be
\phi\equiv \sqrt{\f{3}{2}}M_p\ln\psi\equiv
\kappa^{-1}\ln{\psi}\,,\ee to get a Lagrangian with the usual
kinetic term and a potential $V(\phi)\equiv
M_p^2\mathcal{V}(e^{\kappa \phi})/2$.

We are interested in the situation where there is some {\em local}
fluctuation on a background characterized by some curvature $R_0$
and a density $T_0$. By local we mean that the distance scale of the
fluctuation is short with respect to the characteristic distance
scale associated with the curvature of the background, so that we can
effectively take the flat space limit.
However, as we will see shortly, this does not mean at all that we
can forget about the background curvature $R_0$. In the end, the
reason for this is that the local fluctuations have to match
asymptotic boundary conditions that obviously do depend on the
background.

The weak field expansion now consists of Taylor expanding the
eqs.(\ref{eqhat},\ref{trace}) in powers of the fluctuations and solving
them order by order. So we write:
 \bea
\psi& \equiv &\psi_0(1+ \tilde{\psi})\,,\nonumber\\
\hat{g}_{\mu\nu}&\equiv& \psi_0(g_{\mu\nu}^0+h_{\mu\nu})=\psi_0(\eta_{\mu\nu}+h_{\mu\nu})\,
 \label{deffluct}\,,\\
T_{\mu\nu}&\equiv& T^0_{\mu\nu}+\tilde{T}_{\mu\nu}\,,\nonumber\eea
where the 0 sub/superscript stands for the background value, and as
we said, we take the flat space limit
$g_{\mu\nu}^0\approx\eta_{\mu\nu}$. At first order, the scalar
equation (\ref{trace}) now only depends on the scalar fluctuation
$\tilde{\psi}$ and the trace of the fluctuation of the EMT
$\tilde{T}\equiv \eta^{\mu\nu}\tilde{T}_{\mu\nu}$ \footnote{Since we
are only considering local fluctuations, we can drop terms like
$T_{\mu\nu}^0h^{\mu\nu}$ or $\partial_{\mu}h^{\mu\nu}\partial_\nu
\psi_0$.}: \be (\partial_0^2-\nabla^2)\tilde{\psi}+m_s^2\tilde{\psi}
=-\f{8}{3}\pi G_N^{eff} \tilde{T},\label{eqpsi}\ee where \be
m_s^2\equiv -\f{1}{3}\mathcal{V}_p'(\psi_0)=
\f{1}{3}\left(\f{f'(R_0)}{f''(R_0)}-R_0\right)\,,\ee and we have
defined the effective Newton's constant $G_N^{eff}\equiv
G_N/\psi_0$. Later we will also use the effective Planck mass
defined as: ${M_p^{eff}}^2\equiv \psi_0 M_p^2$.

 Given the matter source $\tilde{T}$ one can now unambiguously
determine $\tilde{\psi}$ at linear order. So let us review here the
case of a spherically symmetric matter distribution with constant density
$\tilde{T}=-\tilde{T}_{00}=-\tilde{\rho}$,  extending over a radius
$r_{\odot}$. The static and spherically symmetric general solution
of (\ref{eqpsi}) outside and inside the distribution reads: \bea
\tilde{\psi}_{out}&=&C_1 \f{e^{-m_sr}}{r}+C_2
\f{e^{m_sr}}{r} \,,\\
\tilde{\psi}_{in}&=&C_3 \f{e^{-m_sr}}{r}+C_4 \f{e^{m_sr}}{r}+\f{8\pi
G_N^{eff} \tilde{\rho}}{3 m_s^2}\,.\eea One then determines the
constants $C_i$ by imposing the appropriate boundary conditions.
First of all we have the condition that $\psi$ takes the background
value at infinity: $\tilde{\psi}(r)\rightarrow 0$ as $r\rightarrow
\infty$. This sets $C_2$ to zero. One then also has the condition
that $\tilde{\psi}$ is regular in the origin, this sets $C_3=-C_4$.
Finally one can determine the other two constants by matching the
solution outside the distribution to the solution inside:
$\psi_{out}(r_{\odot})=\psi_{in}(r_{\odot})$ and
$\psi_{out}'(r_{\odot})=\psi_{in}'(r_{\odot})$. In the case that
$m_s\ll r_{\odot}^{-1}$, one finds in this way
that: \bea \tilde{\psi}_{out}&\simeq& \f{2G_N^{eff}M}{3r}e^{-m_s r}\,,\label{solmssmall}\\
\tilde{\psi}_{in}&\simeq& \f{4\pi G_N^{eff}
\tilde{\rho}}{3}(r_{\odot}^2-\f{r^2}{3})\,,\eea where
$M=4\pi\tilde{\rho}r_{\odot}^3/3$ is the total mass of the
distribution. Notice that when we can neglect the scalar mass term
in (\ref{eqpsi}), even for more general static spherically symmetric
density distributions one can use Gauss's Law to find the same
solution outside the source with $M$ being indeed the total mass of
the distribution. When we can not neglect the scalar mass the
solution outside will depend on the actual density distribution
inside. For instance, in the constant density case, the matching of the
solution outside to the one inside gives for $m_s\gg
r_{\odot}^{-1}$: \bea \tilde{\psi}_{out}&\simeq& \f{4\pi
G_N^{eff}\tilde{\rho}}{3m_s^2}\f{r_{\odot}}{r}e^{m_s(r_{\odot}-r)}\,,\label{solmsbig}\\
\tilde{\psi}_{in}&\stackrel{r\gg m_s^{-1}}{\simeq}& \f{4\pi
G_N^{eff}\tilde{\rho}}{3m_s^2}\left(2-\f{r_{\odot}}{r}e^{m_s(r-r_{\odot})}\right)\,.\eea
We are assuming here that the background is stable so that
$m_s^2\geq 0$. (In fact $m_s^2\sim -R_0$ would also be fine since
for our purpose, the study of local fluctuations, this would mean
$m_s \simeq 0$ effectively.)

We have now found the first order solution for the extra scalar
degree of freedom $\psi$ contained in the metric. The first order
solution for the fluctuations $h_{\mu\nu}$ of $\hat{g}_{\mu\nu}$
follows from the linearization of (\ref{eqhat}) which gives the same
equations as those of Einstein gravity (with a rescaled Newton's
constant): \be \hat{G}_{\mu\nu}^{(1)}=8\pi
G_N^{eff}\tilde{T}_{\mu\nu}\,.\ee So for a static spherically
symmetric mass distribution we find the usual result: \bea
h_{00}&\simeq& \f{2G_N^{eff}M}{r}\,,\\
h_{ij}&\simeq& \delta_{ij}\f{2G_N^{eff}M}{r}\,,\\
h_{0i}&=&0\,.\eea We can now finally write down the linearized
result for the actual metric \be
g_{\mu\nu}=\f{\hat{g}_{\mu\nu}}{\psi}\simeq
\eta_{\mu\nu}+h_{\mu\nu}-\tilde{\psi}\eta_{\mu\nu}. \ee And in the
small scalar mass case, $m_s\ll r_{\odot}^{-1}$, we find for the
metric outside the distribution the familiar result of a
Brans-Dicke scalar-tensor theory with vanishing Brans-Dicke
parameter $\omega$: \bea
ds^2&=&dx^{\mu}dx^{\nu}g_{\mu\nu}\label{BD}\\&\simeq&
-\left(1-\f{2G_N^{eff}M}{r}-\f{2G_N^{eff}M}{3r}e^{-m_s
r}\right)dt^2
+\left(1+\f{2G_N^{eff}M}{r}-\f{2G_N^{eff}M}{3r}e^{-m_s
r}\right)d{\bf x^2}\,.\nonumber\eea

However, this first order result will only be a good approximation
if the weak field expansion makes sense. That is, if the neglected
higher order terms in the equations are subdominant to the
individual first order terms that we have solved for. The weak field
expansion of the kinetic terms in the equations
(\ref{eqhat},\ref{trace}) is the same as for GR, so we already know
that this expansion will only break down as $r_{\odot}$ approaches
the Schwarzschild radius $r_S\sim G_N^{eff}M$, where
$h_{\mu\nu},\tilde{\psi}\sim 1$. The additional ingredient now is
the expansion of the potential term
$\mathcal{V}_p(\psi)\,(\,=-\psi^3\mathcal{V}'(\psi))$ in the scalar
equation (\ref{trace}). Comparing individually the first order terms
that we are keeping
$\psi_0\tilde\psi''(r),\,\psi_0\tilde\psi'(r)/r,\,\psi_0m_s^2\tilde\psi(r)$
 with the higher order terms coming from the Taylor expansion of
 $\mathcal{V}_p(\psi)$ that we are neglecting, we see that the weak
field expansion will be valid as long as \footnote{Throughout this
paper, when writing expressions like $A\gg B$, we will always mean
$|A|\gg|B|$.} \be \f{\psi_0\tilde{\psi}(r)}{r^2}(1+m_s^2r^2)\gg
(\psi_0\tilde\psi(r))^n
\mathcal{V}_p^{(n)}|_{\psi=\psi_0}\,\,\,\,\,\textrm{for
every}\,\,n>1\,.\label{condr0}\ee
 For all the situations that we consider in the next sections,
the terms coming from the second and third derivative of
$\mathcal{V}_p$ will pose no real limitation on the applicability of
the weak field expansion, and the breakdown of the linearization will be
due to the higher order terms, that typically become important at
short distances or high energies. It is then customary to describe the breakdown of the weak field
expansion in terms of a strong coupling scale $\Lambda_s$. One
therefore considers fluctuations characterized by a single energy
scale $E$, such that $h_{\mu\nu},\tilde{\psi}\sim E/M_p^{eff}$ and
$\partial^2h_{\mu\nu},\partial^2\tilde{\psi}\sim E^3/M_p^{eff}$.
The strong coupling scale is then defined as that energy scale
where the weak field expansion breaks down\footnote{This scale is
also relevant when computing for instance quantum corrections, since for
 momenta higher than this the loop expansion will also break down.}. From GR we know that
the expansion of the kinetic terms will be valid as long as $E \ll
M_p^{eff}$. While from the expansion of the potential term in
(\ref{trace}) we now find the strong coupling scale \be
\Lambda_s\equiv\min_{n>3}\left\{M_p^{eff},\left|
\f{1}{\mathcal{V}_p^{(n)}(\psi_0)}\left(\f{M_p^{eff}}{\psi_0}\right)^{n-1}\right|^{\f{1}{n-3}}\right\}\,.
\ee

Notice that this strong coupling scale would be the same as we
would find by simply Taylor expanding the potential of the
canonically normalized field defined in (\ref{action2}) over the
background value $\phi_0 (=\kappa^{-1}\ln\psi_0)$ as \be
V(\phi)\approx \hat{m_s}^2 \tilde{\phi}^2 + X \tilde{\phi}^3 +
\lambda \tilde{\phi}^4 + \sum_{n=0} \frac{
\tilde{\phi}^{4+n}}{(\Lambda_n)^n}. \label{potexp}\ee where
$\tilde{\phi}= \phi-\phi_0$.  The strong coupling scale is then
$\Lambda_s =\psi_0^{1/2}\times\min\{M_p, |\Lambda_n|\}$, where the
factor $\psi_0^{1/2}$ converts the scales in the so called Einstein
frame, described by the metric $\hat{g}_{\mu\nu}\simeq \psi_0
\eta_{\mu\nu}$, to the corresponding scales in the matter frame,
described by the actual metric $g_{\mu\nu}\simeq\eta_{\mu\nu}$.

\section{The spherically symmetric solutions on low curvature backgrounds}

In this section we will focus our attention on $f(R)$ models that
intend to explain the cosmic acceleration through a modification of
gravity at low curvatures. In particular we will apply the results
of the previous section to check under which conditions we can use
the linearization of the action over a background of small curvature
($R_0\sim H_0^2$) to compute the solution that corresponds to a
spherically symmetric mass distribution. As we said we will impose
some conditions on $f(R)$ so that one can talk of the acceleration
as a ``curvature effect'', as opposed to $e.g.$ the $\Lambda$CDM
model that could be described by $f=R-\Lambda$. In particular we
will assume that $f(R)=R + F(R)$ where there is a single mass scale
appearing in $F$ that we will denote by $\mu$ and is of order $\mu
\sim H_0$. Then we will consider separately two cases: the generic
and the fine tuned one. In the generic case we will assume that $F$
is such that \be F^{(n)}|_{R\sim \mu^2} \sim \mu^{2-2n} \sim
H_0^{2-2n}\label{natural} \ee for all $n\geq1$. For $n=1$ we mean by
this that $F'|_{R\sim \mu^2}\sim f'|_{R\sim \mu^2}\sim 1$. For
instance, this is the case of modifications to GR where
$F=\mu^{2n+2}/R^n$ or $F=\mu^2 Log(R)$. We will see that in this
case we can use the linearization in vacuum to study the metric in
all the conventional weak field situations of GR, and the solutions
conflict with observations. In the second case, the fine-tuned one,
we will consider the particular situation in which \be F^{\prime
\prime}|_{R=12H_0^2} \ll \mu^{-2} \sim H_0^{-2}, \ee but still
$F(R)$ satisfies (\ref{natural}) for some $n>2$. This case includes
models like \cite{Dick:2003dw} \be
F(R)=-\f{\mu^4}{R}+\alpha\f{\mu^6}{R^2} \label{tuned1}\,,\ee or like
\cite{Nojiri:2003ft}  \be F(R)=\mu^2 Log(R) + \alpha
\f{R^2}{\mu^2}\,\,\,\,\, {\rm and} \,\,\,\,\, F(R)=\f{\mu^{2n+2}}{R^n} + \alpha
\f{R^2}{\mu^2}, \label{tuned2}\ee with $\alpha$ an order one
parameter, that have been claimed to be compatible with local tests
of gravity. In this case the linearized expansion over the
background with $R=12H_0^2$ breaks down as we approach any source
and therefore can not be used to find the solution. We will
nevertheless see that the true solutions for these models are
essentially the same as in the generic case, with an extra massless
scalar in conflict with observations.  

\subsection{Generic case}

In the previous section we learned that the mass of the scalar and
the range of validity of the weak field expansion follows from the
derivatives of $\mathcal{V}_p(\psi)$ on the background. If the
condition (\ref{natural}) holds one can easily check that $\psi_0$
will be an order one number and that $\mathcal{V}_p(\psi_0)$ and all
its derivatives are of order $\mu^2\sim H_0^2$. We then find that
the scalar is essentially massless $m_s\sim H_0$ and the linearized
solution (\ref{BD}) corresponding to a static spherical mass
distribution is the one of a massless Brans-Dicke scalar field with
vanishing $\omega$ parameter, clearly ruled out by Solar System
tests ($\omega>40.000$ according to the latest measurements from the
Cassini mission \cite{Bertotti:2003rm}). However, as we said, one
should still check if the linearization is applicable, since the
weak field conditions are not necessarily the same as those of GR.
But in this case we find that they actually are: from (\ref{condr0})
we find that the linearization is valid as long as the usual
condition \be \f{G_N M}{r_{\odot}}\ll 1 \ee is satisfied. So we can
use the weak field expansion for the same situations as in GR and in
those cases the solution (\ref{BD}) does indeed describe to a very
high accuracy the static solution that corresponds to a spherical
mass distribution and matches the low curvature background at
infinity. In fact, we see that when translated into an energy scale,
the effects of non-linearities would not be important until we reach
the Planck scale since $\Lambda_s\sim M_p$ , as happens in
conventional GR. There are however significant differences with
respect to GR. Perhaps one of the most striking features of these
solutions is that the scalar curvature remains of order $R\sim
\mu^2$ even inside sources where $G_N T\gg \mu^2$, as also noted in
\cite{Erickcek:2006vf}. But this is easily understood by going to
the definition of the field $\psi$. For instance, in the case when
$F=\mu^{2n+2}/R^n$ we have that \be \psi=1-n\f{\mu^{2n+2}}{R^{n+1}}.
\ee so for $R\sim \mu^2$ a shift in the scalar curvature of order
$\mu^2$ implies a shift in $\psi$ of order one which corresponds to
an energy scale of order $M_p$ for the canonically normalized field
$\phi$. In these models, once the background scalar curvature
reaches its vacuum value it stays locked into this $R\sim \mu^2$
regime, and we would need extremely energetic processes involving
strong fields ($\tilde{\phi} \geq M_p$) to get it out of this
``locked phase'' locally.

\subsection{Fine-tuned case}

In \cite{Dick:2003dw,Nojiri:2003ft} several models are proposed for
which the second derivative of $F$ is zero on a particular low
curvature (de Sitter) background. It is argued that since the mass
of the scalar $m_s^2\sim 1/F''(R)$ now goes to infinity, its action
range goes to zero and the model becomes compatible with the Solar
System experiments. In fact, to be compatible with the current
limits from fifth force search experiments, $m_s\geq 10^{-3}eV$
would already be fine \cite{Adelberger:2003zx}. For instance in the
case (\ref{tuned1}) one has \be m_s^2 =-
\f{R_0^2}{2}\f{(1-\f{8\mu^2}{3R_0})}{(R_0-3\alpha
\mu^2)}\,\label{formmas}\ee and one finds that $m_s\geq 10^{-3}eV$
for \be R_0=3\alpha\mu^2(1 \pm 10^{-60}) \,.\ee This already shows
the problem with this proposal. If the background curvature deviates
only minutely from this fine-tuned value, the scalar would become
light again, in conflict with experiment, as we already noticed in
\cite{Navarro:2005ux}. However, one might still think that for these
particular fine tuned backgrounds the static solution corresponding
to a spherically symmetric mass distribution would be very close to
the GR solution, corresponding to (\ref{BD}) with large $m_s$. But,
just as in the previous subsection, we should check the weak field
conditions to make sure that we can actually use the linearized
solution (\ref{BD}). Let us therefore consider the form of
$\mathcal{V}_p(\psi)$ in the neighbourhood of the background for
which $F''=0$ (and where $m_s^2\rightarrow \infty$). We will denote
by $R_t$ and $\psi_t$ the values of $R$ and $\psi$ in such
background. As we said, even on this fine-tuned background the
condition (\ref{natural}) will still be satisfied for some $n>2$.
For instance in the cases (\ref{tuned1},\ref{tuned2}) this condition
 is satisfied for all $n>2$. In particular we have
$F'''(R_t)\sim 1/\mu^4$ and one can show from (\ref{defpsi},\ref{defpsi2}) and
(\ref{defF}) that:
\bea \psi-\psi_t&\simeq& F'''(R_t)\f{(R-R_t)^2}{2}\,,\\
\nonumber\\
\mathcal{V}_p(\psi)-\mathcal{V}_p(\psi_t)&\simeq&-f'(\mathcal{R}(\psi_t))(\mathcal{R}(\psi)-\mathcal{R}(\psi_t))\nonumber\\
&\simeq&\pm\sqrt{2}\psi_t
\left(\f{\psi-\psi_t}{F'''(R_t)}\right)^{1/2}\,,\label{branch}\eea
where the approximations are valid for $(R-R_t)\ll\mu^2$ or
equivalently $(\psi-\psi_t)\ll 1$ and the sign in (\ref{branch})
depends on the chosen branch for $\mathcal{R}$. Looking at the
approximate form of $\mathcal{V}_p$ for $\psi$ close to $\psi_t$ it
is obvious why the Taylor expansion over a background with
$\psi_0=\psi_t$ breaks down, since $\mathcal{V}_p$ is non-analytic
at this point. This should not come as a surprise, because we know
that when $F''$ becomes zero one can not invert the relation
$f^\prime(R)=\psi$ and $\mathcal{R}$ gets a branch
point. In any case, one can see that the weak field conditions
(\ref{condr0}) for the expansion on a background with curvature
$R_0$ close to $R_t$ now simply reduce to the condition for the
validity of the expansion of $\mathcal{V}_p(\psi_0(1+\tilde{\psi}))$
in powers of $\tilde{\psi}$, which are \be \tilde{\psi}\ll
\psi_0-\psi_t\sim \f{1}{F'''(R_t)\mathcal{V}_p'(\psi_0)^2}\sim
\left(\f{\mu}{m_s}\right)^4\sim \left(\f{H_0}{m_s}\right)^4\,. \ee
Here $m_s$ corresponds to the mass of the scalar in the background
with $\psi=\psi_0$. Plugging here the expressions for the linearized
solutions for $\tilde{\psi}$ we find that this condition becomes \be
\f{G_NM}{r_{\odot}}\ll \f{H_0^4}{m_s^4}\,,\ee for objects with size
$r_{\odot} \ll m_s^{-1}\,,$ and \be G_N \tilde{\rho}\ll
\f{H_0^4}{m_s^2}\,,\ee for objects with size $r_{\odot}\gg
m_s^{-1}$.  And it is apparent that if we take $m_s\geq 10^{-3}eV$
we can not use the linearized solution in any situation. This
failure of the weak field expansion is also manifested in the very
low strong coupling scale that we now find: \be \Lambda_s\sim M_p
\left(\f{H_0}{m_s}\right)^4\,,\ee which is even smaller than the
Hubble scale $(\leq H_0^2/M_p)$ for $m_s\geq 10^{-3}eV$.

To find the true static solution corresponding to a spherically
symmetric mass distribution with density $\tilde{\rho}$ we should
look at the full equation (\ref{trace}). Let's take for instance the
specific background with curvature $R_0=R_t$. If this static
background is solution of the EOM we have that
$\mathcal{V}_p(\psi_t)=0$, so using the approximate expression for
$\mathcal{V}_p$, eq.(\ref{branch}), the scalar equation
(\ref{trace}) now becomes: \be \nabla^2\tilde{\psi}\,\, \pm
\left(\f{2\psi_t\tilde{\psi}}{9F'''(R_t)}\right)^{1/2} =-\f{8}{3}\pi
G_N^{eff} \tilde{\rho}.\label{exact}\ee Notice that this is valid as
long as $\tilde{\psi},h_{\mu\nu}\ll1$. The correct approximate form
of the equation does not correspond to a Klein-Gordon equation with
a large mass term, but an unusual power $\tilde{\psi}^{1/2}$ of the
fluctuation appears in the potential term. Had we inappropriately
used a Taylor expansion of the potential we would have erroneously
concluded that the mass term appearing in the equation diverges.
Imagine now that we have a solution of this equation that matches
the asymptotic background at infinity, so $\tilde{\psi}\rightarrow
0$ for $r\rightarrow \infty$ and corresponds to the spherical source
with mass density $\tilde{\rho}$. Keeping the correct approximate
form of the equation we can see that the Laplacian will dominate
over the potential term at short distances as usual and the solution
close enough to the source will be the one of a massless scalar to a
good approximation: \be \tilde{\psi}=\f{2G_N^{eff}M}{3r}+C\,. \ee We
have not fixed the constant $C$ because its precise value would
depend on the non-linear interactions of $\psi$ that will be
important at long distances. By comparing the terms in the Laplacian
that we have solved for, with the potential term that we have
ignored, we see that the solution corresponding to a massless scalar
will be a good approximation as long as $r\ll (G_NM/\mu^4)^{1/5}$ if
$C<(\mu G_N M)^{4/5}$ or as long as $r\ll (G_N
M/(\sqrt{C}\mu^2))^{1/3}$ if $C>(\mu G_N M)^{4/5}$. So we see that
regardless of the precise value of $C$ (that, as we said, should be
determined by matching with the asymptotic background), the solution
to the full equation (\ref{exact}) in this fine-tuned case is the
same as the one we found in the generic case for small distances.

Another specific problem with models of the type (\ref{tuned2}),
that include terms like $\sim R^2$ in the action, is that one does not even recover
Einstein gravity for high curvature backgrounds. One can see this
immediately from the Lagrangian because it can be approximated by
$f\simeq R+\alpha R^2/\mu^2$ when $R\gg \mu^2$. In terms of the
extra scalar fluctuation this means $m_s^2\sim R_0$ for large
background curvatures $R_0\gg \mu^2$. Notice that now
$\Lambda_s\sim M_p^{eff}$, so that we can indeed trust the
linearization, and we would find an extra long range force when
performing local experiments.  In the next section we will show how
the extra scalar fluctuation disappears for large background
curvatures, if the action {\em does} approximate the
Einstein-Hilbert action, $f\simeq R$ for $R\gg\mu^2$.

Finally, we have only explicitly considered the case with
$F'''(R_0)$ equal to its 'natural' value. One could imagine an even
more fine-tuned case where $F^{(n)}(R_0)\ll H_0^{2-2n}$ for all
$n<m$ and (\ref{natural}) is satisfied only for $n\geq m$. In that case one finds for $\psi-\psi_t\ll1$: \be
\mathcal{V}_p(\psi)-\mathcal{V}_p(\psi_t)\sim
\mu^2(\psi-\psi_t)^{1/(m-1)}\,,\ee resulting in a strong coupling
scale \be \Lambda_s\sim
M_p\left(\f{H_0}{m_s}\right)^{\f{2m-2}{m-2}},\ee so we see that in
any case $\Lambda_s\leq M_p\left(\f{H_0}{m_s}\right)^2$ ($\sim H_0$
for $m_s\sim 10^{-3}eV$). The correct approximate
equation for the scalar, eq.(\ref{exact}), would now generalize to
\be
\nabla^2\tilde{\psi}\,\, + \f{1}{3}\left(\f{(m-1)!\psi_t\tilde{\psi}}{F^{(n)}(R_t)}\right)^{1/(m-1)} =-\f{8}{3}\pi G_N^{eff}
\tilde{\rho},\ee
and the same argument can be used to show that the solutions will
approach those of the scalar-tensor theory with a massless scalar at
short distances.

\section{The spherically symmetric solutions on high curvature backgrounds}

\subsection{The Chameleon effect}

Let us begin this section by briefly reviewing the Chameleon
mechanism \cite{Khoury:2003aq} for hiding the effects of a field
that is otherwise very light in vacuum. Usually one assigns a range
to the force mediated by a given field according to its mass, because when $r> m^{-1}$ the potential produced by the source gets an
exponential Yukawa suppression. However, when non-linear
interactions are important there are more possibilities. Imagine
that we have a scalar field with an arbitrary potential in vacuum,
$V(\phi)$, and a coupling to the trace of the matter EMT
like\footnote{$T\leq 0$ for all realistic cases, so it is convenient
to define $\rho\equiv -T\geq0$.} $\Delta {\cal
L}=\alpha(\phi)T=-\alpha({\phi})\rho$. For finding static solutions
we should solve the equation \be \nabla^2 \phi =
\alpha^\prime(\phi)\rho + V^\prime(\phi).\label{eqstatic} \ee Notice
that the solution of this equation is unique for a given source and
asymptotic boundary conditions. To understand the Chameleon effect
we will consider two different spherically symmetric situations: in
one $\rho=0$ everywhere except in a small region $r\leq r_\odot$
where it is constant, while in the other $\rho$ is a constant
everywhere.

Let's deal now with the first case. If we can use a weak field
expansion the solution will be
completely analogous to the solution of the linearized EOM of
$f(R)$ gravity presented in the second section but let us briefly
repeat it here. The asymptotic value of the field, $\phi_0$, is
such that it minimizes its vacuum potential, $V^\prime(\phi_0)=0$.
Then in the region where $\rho\neq 0$ the field finds itself in a
non-equilibrium position, and it will acquire a non-trivial
profile. In a weak field expansion in powers of the fluctuation
$\tilde{\phi} = \phi-\phi_0$, the linearized equation
becomes  
\be \nabla^2 \phi = \frac{\beta \rho}{M_p} + m_s^2(\phi-\phi_0),\label{lineq} \ee
where $m_s^2\equiv V^{\prime\prime}(\phi_0)$ and $\beta \equiv M_p
\alpha^\prime(\phi_0)$. The solution outside the source ($r\geq
r_\odot$) is \be \phi(r) = \phi_0 + C_1 \frac{e^{-m_s r}}{r},
\label{linsol}\ee where we have taken into account the asymptotic
boundary conditions and $C_1$ is a constant to be determined. The
solution inside the source is \be \phi(r) = \phi_0 -\frac{\beta
\rho}{M_p m_s^2} + C_2 \frac{e^{-m_s r}}{r} +C_3 \frac{e^{m_s
r}}{r}. \ee Now we can determine the constants $C_i$ by imposing
$\phi^\prime(0)=0$ and continuity of $\phi$ and its first derivative
at $r=r_\odot$. Doing this, in the limit $r_\odot \ll m_s^{-1}$ we
get the usual result \be C_1 = -\frac{\beta \rho r_\odot^3}{3 M_p}
=- \frac{\beta M}{4 \pi M_p}, \label{linconst}\ee where $M\equiv
\frac{4}{3}\pi r_\odot^3 \rho$. We see how the localized
energy-momentum density sources the field and makes it acquire a
non-trivial profile outside the source.

However, in the second situation where $\rho$ is a constant
everywhere, this term can be seen as another contribution to the
potential of $\phi$. So the field will not acquire a non-trivial
profile but will simply set to the minimum of its $\rho$-dependent
effective potential $V_{eff} = V(\phi) + \alpha(\phi)\rho $. So the
solution will be $\phi=\phi_s$ where $\phi_s$ satisfies
$V^\prime(\phi_s) +\alpha^\prime(\phi_s)\rho=0$.

The Chameleon effect will take place whenever the second situation
is a good approximation for the solution inside a localized
spherically symmetric source. When this is the case, the region
inside the source where the field is settled to the minimum of its
effective potential will not source the field outside. The ``thin
shell'' will be the region near the surface of the massive body
where the field has a profile interpolating between its equilibrium
positions, $\phi_0$ and $\phi_s$. If this region has a thickness
given by $\Delta r$, the effects of the force mediated by this field
will be suppressed as long as $\Delta r/r_\odot \ll 1$. For the
thickness of this ``thin shell" one finds the approximate expression
\cite{Khoury:2003aq}: \be \frac{\Delta r}{r_\odot}\simeq
\frac{(\phi_0 - \phi_s)}{6\beta \Phi_N M_p}\label{ts}, \ee where
$\Phi_N\equiv G_N M/r_{\odot}$ is the Newtonian potential at the
surface of the body. So the condition for the Chameleon mechanism to
hold will be that the difference of the values of the field that
minimize the effective potential inside and outside the source
should be much smaller than $\beta \Phi_N M_p$. When non-linearities
are negligible we can approximate the potential and the coupling
function by $V=m_s^2 \phi^2$ and $\alpha= \beta_0 \phi/M_p$. It is
easy then to compute $\phi_s$ and applying the previous formula we
see that there will be a thin shell only when $r_{\odot}\gg
m_s^{-1}$, and the effects of the field will be hidden only for
distances larger than the inverse mass, as expected. But when
non-linearities are relevant we can have a Chameleon effect even if
$r_{\odot}\ll m_s^{-1}$. Notice that this necessarily implies the
breakdown of the weak field expansion. In this case only the mass
contained within the thin shell will contribute to the field outside
the source so we can estimate the solution for the scalar field
outside the source as \cite{Khoury:2003aq} \be
 \phi \simeq \phi_0 -\frac{\beta M_{ts}}{4 \pi M_p}\frac{e^{-m_s r}}{r} \simeq \phi_0 -\frac{3\Delta r}{r_\odot}\frac{\beta M}{4 \pi M_p}\frac{e^{-m_s r}}{r},\label{camsol}
\ee where $M_{ts}$ is the mass contained within the thin shell.
Through this mechanism a field that is very light in vacuum could
have passed experimentally undetected.

\subsection{The Chameleon effect in $f(R)$ gravity}

We are now in a position to assess what are the necessary properties
for an $f(R)$ action in order to get a Chameleon effect for the
extra scalar degree of freedom. To parallel the Chameleon literature
\cite{Khoury:2003aq} we will work with the canonically normalized
field $\phi$ defined in eq. (1). For $f(R)$ theories one can see
from the action (\ref{action2}) that the coupling of the scalar
field to matter is given by $\Delta {\cal L}= e^{-2\kappa
\phi}T/4\equiv- e^{-2\kappa \phi}\rho/4$ where $\kappa^{-1}=
\sqrt{3/2}M_p$ and $T$ is again the trace of the EMT. Qualitatively,
what we need to get a Chameleon effect is that as $\rho$ becomes
large, the value of $\phi$ that minimizes the effective potential,
$V_{eff}=V(\phi)+e^{-2\kappa \phi}\rho/4$, depends very weakly on
$\rho$. If this is the case the difference $\phi_0-\phi_s$ in the
estimation of the thin shell thickness of the previous section will be small when the asymptotic
background energy density is large and massive bodies will indeed develop a
thin shell, so that the extra force becomes negligible. When this is
the case the scalar curvature will follow roughly the Einstein equations, $R\simeq \rho/M_p^2$. The scalar curvature is however given
in terms of $\phi$ trough its definition eq.(\ref{inv}), so the
requirement that as $\rho$ becomes large the variation of the
equilibrium value for $\phi$
becomes small can be rephrased as the requirement that as $R$
becomes large, $f^\prime(R)$ becomes roughly constant. So, for
modifications of the GR action characterized by a curvature scale
$\mu^2$, we will have a Chameleon effect at high values of the
background energy density, $\rho_0\gg \mu^2 M_p^2$, (or curvature
$R_0 \gg \mu^2$) if \be f^\prime(R) \simeq 1 \;\;\;\;\; {\rm when}
\;\;\;\;\; R\gg \mu^2. \label{condhigh}\ee This qualitative
discussion agrees with the naive expectations that one could have,
since we are simply saying that the effects of the extra degree of
freedom will be hidden when the background curvature is large if the
action $looks$ $like$ the Einstein-Hilbert action when the curvature
is large. However, in order to consider these large curvature
backgrounds we need them to be relatively stable. This will give a
condition on the departure from the Einstein-Hilbert action for
large curvatures. From (\ref{condhigh}) we see that there are
essentially two possibilities. Either $f''(R)$ will be close to zero
and positive; or $f''(R)$ will be close to zero but negative. In the
latter case we find a large tachyonic instability for the scalar
fluctuation ($m_s^2\sim 1/f''(R)$), resulting in a decay of the
large curvature background. For instance, the models
$f(R)=1-\mu^{2+2n}/R^n$ that were proposed
originally \cite{Carroll:2003wy}, have $m_s^2\sim -R (R/\mu^2)^{n+1}$
for large curvatures, so for such models those backgrounds are
unstable. For this reason they fail to give a realistic expansion of
the Universe at early times, as noticed in \cite{Song:2006ej}. This
also agrees with the results of \cite{Amendola:2006kh}, where it was
shown that FRW solutions for these models never attain large
curvatures, clearly in conflict with a conventional matter dominated
expansion.

The story is completely different for the models with positive
values of $f''(R)$ at large curvatures, that we will consider from
now on.  The scalar fluctuation now has a large positive mass
squared and the backgrounds are stable. So we can safely assume that
the FRW solutions will only start to differ from those of Einstein
gravity at the current epoch, when $H\sim H_0\sim \mu$. One might
think that the recovery of Einstein gravity for large background
curvatures now simply happens because the mass becomes large,
reducing the range of the extra force. However, this assumes that
the weak field expansion remains valid. But, as we preempted in the
beginning of this subsection, the weak field expansion will in fact
break down for realistic situations, and the recovery of Einstein
gravity happens through the non-linear Chameleon mechanism. For a
massive body of mass $M$, with radius $r_{\odot}$ and density $\rho
=3M/(4\pi r_{\odot}^3)$ we then get the following picture. For low
values of the asymptotic background energy density $\rho_0$ the
gravitational field of the body will be of the scalar-tensor type,
with an extra force with long range $1/m_s$. If we now increase the
background density and curvature, the range of the extra force will decrease. But
then at some point, when $1/m_s$ is still larger than $r_{\odot}$,
the linearized solution breaks down and the body develops a thin
shell, suppressing the extra force. Let us now illustrate this for
some particular form of the function $f$. We will consider functions
such that when $R\gg \mu^2$ can be approximated by\footnote{Notice
that we are not making any assumption here about the
  properties of the vacuum solutions (where $R\sim \mu^2$) since we
  are just assuming a
  functional form for $f$ that holds in the high curvature limit $R\gg
\mu^2$.} $f(R)\simeq R + \mu^{2n+2}/R^n$. The relation of the scalar
field with the curvature will then be given by \be e^{\kappa \phi}
\simeq 1-n\frac{\mu^{2n+2}}{R^{n+1}}. \ee Since we are assuming that
$R$ is positive and much bigger than $\mu$ we will be interested in
the behavior of the potential for negative values of $\phi$ close to
zero. Our assumptions imply then that in this region the effective
potential for the field can be approximated by \bea V_{eff} &\simeq &
e^{-2\kappa \phi}\left(\f{\rho_0}{4}-\f{n+1}{2}M_p^2\mu^2
\left(\f{1-e^{\kappa
\phi}}{n}\right)^{\frac{n}{n+1}}\right)\\&\simeq&\f{\rho_0}{4}e^{-2\kappa
\phi}-\f{n+1}{2}M_p^2\mu^2(-\f{\kappa\phi}{n})^{\f{n}{n+1}}
\label{pot}. \eea For $\rho_0\gg M_p^2\mu^2$ this potential is
minimized for \be \kappa\phi_0 \simeq -n\left(\frac{M_p^2
\mu^2}{\rho_0}\right)^{n+1}. \ee We see how the equilibrium value of
the field simply gets closer to zero as we increase the background
energy density. As we just said, this will give a large mass for the
scalar fluctuations on these large curvature backgrounds: \be
m_s^2=V_{eff}''(\phi_0)\sim \mu^2(-\kappa
\phi_0)^{-\f{n+2}{n+1}}\sim\f{\rho_0}{M_p^2}\left(\f{\rho_0}{M_p^2\mu^2}\right)^{n+1}=
R_0\left(\f{R_0}{\mu^2}\right)^{n+1}\,.\label{mass}\ee However, at
the same time this gives a low strong coupling scale for the weak
field expansion. Indeed, the expansion of the potential (\ref{pot})
in powers of the fluctuation $\tilde{\phi}$ breaks down for
$|\tilde{\phi}|\geq |\phi_0|$, so we get a strong coupling scale \be
\Lambda_s\sim |\phi_0|\sim M_p\left(\frac{M_p^2
\mu^2}{\rho_0}\right)^{n+1} \,, \ee and the linearized solution for
$\tilde{\phi}$, that we get from eq.(\ref{lineq}) with
$\beta=-\kappa M_p/2$, is only valid when \be
\kappa\tilde\phi(r_{\odot})\simeq \f{2G_N M}{3r_{\odot}}\ll |\kappa
\phi_0|\sim\left(\frac{M_p^2 \mu^2}{\rho_0}\right)^{n+1}\,.
\label{breakdown} \ee For a given source, this will be the case for background densities
$\rho_0$ smaller than a certain critical density $\rho_c$ given
by\footnote{We are assuming here that the density of the
  body, $\rho$, is $\rho\gg \rho_c$; one can
show from (\ref{mass}) that this indeed implies that
$1/m_s(\rho_c)\gg r_{\odot}$.} \be \rho_c\equiv\f{
M_p^2\mu^2}{\Phi_N^{1/(n+1)}}\,,\ee  where $\Phi_N$ stands again for
the Newtonian potential at the surface of the body. If the
background density is much larger than $\rho_c$ the linearization
breaks down and the full non-linear equation is approximately solved
by the Chameleon thin shell solution (\ref{camsol}). Indeed, from
eq.(\ref{ts}) we find that the body develops a thin shell if: \be
\frac{\Delta r}{r_\odot}\sim \left(\frac{M_p^2
\mu^2}{\rho_0}\right)^{n+1}\Phi_N^{-1}=
\left(\f{\rho_c}{\rho_0}\right)^{n+1} \ll 1\label{thinshell}. \ee
Notice that massive bodies with strong gravitational fields in
their surface ($i.e.$ large $\Phi_N$) will develop a thin shell more
easily than smaller ones for a given asymptotic background
curvature. We also see how, when we take the limit $\mu \rightarrow 0$
leaving $\rho_0$ fixed, all sources will develop an (infinitely)
thin shell and we recover the solutions of GR, in which the field
(curvature) is a constant outside the source and jumps
discontinuously ($\Delta r \rightarrow 0$) to a different value
inside the source. So we see that the true limit to GR in these
models would be $\f{\mu^2M_p^2}{\rho_0}\rightarrow 0$, where
$\rho_0$ is a background energy density.

In these estimations, if we are dealing with the cosmological
background, we can approximate $\frac{\rho_0}{M_p^2 \mu^2}\sim
(1+z)^{3}$. Assuming then an adiabatic evolution ($i.e$ the solution
is always taken to be the equilibrium, static one for the prescribed
asymptotic value of the curvature,) we can estimate using
eq.(\ref{thinshell}) at what cosmological time a given isolated
source would change its gravitational field from GR to
scalar-tensor. For instance a star has typically $\Phi_N \sim
10^{-6}$, while a galaxy or galaxy cluster can have $\Phi_N$ in the
range $\sim 10^{-4}- 10^{-7}$. It is clear thus from (\ref{thinshell})
that we can expect
that the gravitational field of these astrophysical sources would
have changed to the scalar-tensor type already at high redshifts,
when $(1+z)\sim 10^{4/(3n+3)}-10^{7/(3n+3)}$. Notice that if we had
used the invalid linearized result (\ref{linsol},\ref{linconst}), we would simply
assign a range to the force mediated by the field given by its
inverse mass, $r_s \sim m_s^{-1}$, and we could estimate the
redshift dependence of this distance as $r_s(z) \sim m_s^{-1}(z)
\sim \mu^{-1}(1+z)^{-3(n+2)/2}$. So we would conclude that the field
would have very long range even at high redshifts. However, here we
have shown that non-linear interactions provide a further
suppression of the effects of the scalar field through the Chameleon
effect.

We should point out here that in this section we have just presented
order-of-magnitude estimations for the necessary conditions to
recover Einstein gravity. But since these solutions are
non-perturbative, in order to study its behavior at a more
quantitative level when for instance $\Delta r/r_\odot$ is not very
small, a numerical integration of the equations would be mandatory.
Also, we have assumed an adiabatic evolution in the estimation of
the cosmological time at which the gravitational field of a given
source would change from GR to scalar-tensor. Under this assumption
the models would be ruled out, since we would be in the scalar-tensor
regime at present. But to study the time-dependence of
these solutions, thereby checking explicitly if their evolution is
really adiabatic, one might have to resort again to numerical
methods. On the basis of our analysis we can not exclude that models
exist for which the GR to scalar-tensor transition is slow
and non-adiabatic. In this case such $f(R)$ models could perhaps be brought into
accord with observations.

\section{Conclusions}

In this paper we have studied the solutions corresponding to
spherically symmetric sources in $f(R)$ theories of gravity. We have
started by clarifying the range of validity of the linearized
expansion, giving the conditions that have to be satisfied for this
expansion to be valid for an arbitrary function $f$. Then we have shown that for the models that represent a
modification of gravitational dynamics only at low curvatures, the
linearized expansion on vacuum breaks down as long as the scalar
curvature departs significantly from its vacuum value. These models
are characterized by a function that can be approximated by the
Einstein-Hilbert action ($f\simeq R$) when $R \gg \mu^2$, but is
non-trivial when $R\sim \mu^2$. However, when we fix our asymptotic boundary conditions for
the curvature to the vacuum value, in most cases the linearized
solutions (that are in clear conflict with Solar system experiments)
are valid as long as $G_NM/r \ll 1$, the same condition that one
finds in GR. Indeed, the strong coupling scale of this linearization
is the Planck mass, as in GR. This implies that, in these models, once the scalar
curvature diminishes to a value $\sim \mu^2$, it gets ``locked''
into this value everywhere, and one would require very strong
gravitational fields or Planck scale energies to be able to get out
of this phase locally. It is worth to mention here that the
situation is completely different for models that include inverse
powers or logarithms of other curvature invariants beyond the scalar
curvature. In particular if we include invariants that do not vanish
in the Schwarzschild solution, even in the case when the linearized
solutions are the same, the strong coupling scale for the
linearization on vacuum is significantly smaller, $\Lambda_s\sim
(\mu^3M_p)^{1/4}$, as we showed in \cite{Navarro:2005gh}. For those
models one can never use the linearized solutions, and one does
recover the GR solutions at short distances as required by Solar
System or laboratory experiments \cite{Navarro:2005gh}.

Also, we have seen why raising the mass of the scalar field in
vacuum by adding, $e.g.$ an $R^2$ term to the action does not imply
compatibility with local tests of gravity, as claimed in
\cite{Nojiri:2003ft}. In fact, it is easy to see that those models
are ruled out because when adding positive powers of $R$ to the
action we are modifying gravity also at high curvatures. For
instance in the $R^2$ case one can approximate the action by $f=R +
R^2/m^2$ with $m\sim \mu$ in the high curvature regime ($R\gg
\mu^2$). In those situations the scalar field becomes effectively
massless and its effects would have been observed. Furthermore, we
have argued that even the solutions in vacuum are of scalar-tensor
type for observationally relevant distances in these cases.

On the other hand, for the models that modify gravity only at small
curvatures, when the asymptotic value of the curvature is large
($i.e.$ $\gg \mu^2$ because there is a large, constant background
energy-momentum density), the linearization breaks down and the
behavior of the extra scalar field, that we can associate to the
extra degree of freedom contained in the metric, is governed by
non-linear dynamics. This observation allows one to recover
quantitatively what intuitively could seem obvious: the solutions of
the theory approach those of GR when $R\gg \mu^2$. This is achieved
through the so-called Chameleon mechanism \cite{Khoury:2003aq}:
whenever there is a localized massive body in a high curvature
background, the scalar field quickly adopts a new equilibrium position
inside it as a consequence of its non-linear interactions even if
its mass is small on the background. As a consequence only a ``thin
shell'' of matter in the surface of the body acts as a source for
the field outside. This ``thin shell'' is the region near the
surface where the scalar field interpolates between its equilibrium
positions. We have estimated this thickness for some forms of the
function $f$, giving a quantitative estimation of the necessary
value of the background curvature in order for this effect to take
place for a body of given mass and radius. Using this estimation,
and under the assumption of an adiabatic evolution, we have argued
that in the application of these theories to the Universe one should
observe a transition from GR to scalar-tensor gravity as the
Universe expands, taking place already at high redshifts for most
astrophysical sources. However we have not ruled out the possibility
that this gravitational decay could be slow and non-adiabatic for
some models. If the Earth and its environment would still remain in
the GR regime in some cases, those models would be compatible with
local tests of gravity. But a
 quantitative determination of the time scale associated to the decay of
the scalar curvature to its equilibrium value $\sim \mu^2$ inside
sources as the asymptotic background curvature approaches its vacuum
value would require the study of time-dependent non-perturbative
solutions which lie beyond the scope of the present paper.

Finally, our analysis also allows us to make some more general
statements on the applicability of the linearization for these
theories. For low curvatures of the background $R_0\sim \mu^2$ one
can safely use the weak field expansion in the same situations as
for GR. However, for large curvatures, one has to be careful when
using this expansion. For instance, the linearized equations for the
cosmological perturbations that were used in \cite{Song:2006ej}
will become invalid at early times, at least if one considers
realistic ($i.e$ high curvature) FRW backgrounds. The recovery of Einstein gravity is
essentially non-perturbative as we illustrated in the previous
section. And in such models we can expect that in general there
will be a range of background curvatures for which the linear
solutions are those of scalar-tensor gravity, whereas the true
non-linear solutions are in agreement with GR.

\section*{Acknowledgements}
We acknowledge discussions with Eanna Flanagan and Luca Amendola.
K.V.A. is supported by a postdoctoral grant of the Fund for
Scientific Research-Flanders (Belgium).

\end{document}